# MEMS Sensor for Detection and Measurement of Ultra-Fine Particles: A Review

Vinayak Pachkawade[a,b*] and Zion Tse[c]

[a*] SenseAll, India

[b] Independent Researcher, India

[c] Department of Electronic Engineering, University of York, Heslington, York YO10 5DD, United Kingdom

*Contact details of the corresponding author

Vinayak Pachkawade

Email: vinayakp@ieee.org

Phone: +91-89995 82346

ORCID: vinayak pachkawade

https://orcid.org/0000-0001-7357-3350

**Abstract**

This paper investigates the performance of the micro-electro-mechanical systems resonant sensor used for particle detection and concentration measurement. The fine and ultra-fine particles such as particulate matter (PM), ferrous particles, and nanoparticles are known to contaminate the atmosphere, fluids used in industrial machines, and food, respectively. The physical principles involved in the target particles accumulating on the sensor are presented. Micro-gravimetric resonators that use piezoelectric and thermally actuated transducers for particle detection and concentration measurement in air and high-viscosity liquids are analysed. Critical sensor features, such as maximum possible parametric sensitivity, the detection limit of particle size and mass concentration, linear dynamic range, and output stability, are thoroughly evaluated.

## 1. Introduction:

Environmental conditions have had a worldwide, major, and long-lasting impact on our health, productivity, and overall well-being. The World Health Organization (WHO) states that by reducing air pollution levels, countries can decrease the burden of various diseases, such as heart disease, stroke, lung cancer, and both chronic and acute respiratory diseases, including asthma [1]. As per the latest report, a very large percentage of the world population is living in places where the WHO air quality reference levels are not met [1–3]. Outdoor air pollution is estimated to cause millions of early deaths worldwide in the report. Similarly, indoor pollution (e.g., smoke) is considered a serious health risk for billions of people. Lower levels of air pollution are linked with better long-and short-term cardiovascular and respiratory health. To address these challenging global health problems, various policies are being improved to reduce indoor and outdoor air pollution [4]. From a technological point of view, environmental parameter sensing opens up a range of new possibilities to create smarter devices to advance human health and well-being [5].

Currently, a wide range of sensing solutions are being developed to provide detailed and reliable data on key environmental parameters such as particulate matter (PM) [6–10]. Some specific examples of these parameters are $PM_{2.5}$, which refers to PM with particle diameter less than or equal to 2.5 μm, and $PM_{10}$, which refers to PM with particle diameter less than or equal to 10 μm. High concentrations of PM in the surrounding environment indicate the presence of bacteria, viruses, and/or other harmful elements that may cause direct and profound impacts on our health and well-being [11]. The revised guidelines set by the WHO define the limits on the amount of PM that a human body can tolerate without risking respiratory or cardiovascular diseases. These limits include an annual mean of 5 μg/m$^3$ and 24-hour mean of 15 μg/m$^3$ for $PM_{2.5}$, and an annual mean of 15 μg/m$^3$ and 24-hour mean of 45 μg/m$^3$ for $PM_{10}$ [12]. $PM_{2.5}$ is considered one of the most dangerous air pollutants. Due to their size, $PM_{2.5}$ particles can travel deep into human lungs and cause a range of health problems. It is estimated that about 15% of the worldwide deaths from the ongoing coronavirus disease 2019 (COVID-19) pandemic could be linked to long-lasting exposure to air polluted by anthropogenic sources (e.g., PM from combustion sources) [13]. Ultra-small particles, such as $PM_1$, ferrous debris, and other nanoparticles, are considered even more harmful for humans, food, and industrial machines [7, 14–19]. Therefore, the detection and measurement of ultra-small particles in media such as the surrounding atmosphere and various liquids are most important to protect human health, food quality, and industrial machine maintenance.

## 2. Theory

There are two main types of instruments available to monitor fine and ultra-fine particles in the surrounding air. The first type of instrument is based on the gravimetric method of directly measuring the mass of particles [20]. Particles are collected on a filter over a fixed period of time and are then weighed in a laboratory. Such methods are expensive, and the required instrumentation size restricts its wide usage (e.g., deploying the sensor network) [20]. The second type of instrument is based on the principle of laser scattering [21]. In this category, particles (in air or fluid) are illuminated with light of a certain wavelength. The amount of the scattered light provides an approximation of the number of particles in the environment. These sensors rely on several assumptions to estimate the density and size distribution of particles, thus leading to inaccurate results. Moreover, these sensors use sophisticated optical elements

[22][23], are complicated to use, and are also relatively expensive [21]. A third type of instrument involves counting particles by measuring the current of charged particles [18], but such instruments are typically quite large. The challenges linked with PM sensors based on electrochemical technology [24], metal oxides [25], etc. are various, including that they are bulky, consume relatively more power, and have poor long-term reliability.

Commonly used tests to determine the stability of industrial oil can determine the oxidative resistance of oils [15, 16, 26]. However, these tests are highly time consuming and expensive, thus limiting their applicability to regularly scheduled oil checks. The viscosity and acidity of oil can be measured using sensors integrated in lubricated machinery [27]. Such methods can monitor changes of parameters (viscosity, density, oil resistance, oil acidity) that could lead to the formation of particles. However, none of these tests can directly detect particles, allowing for the detection of even subtle changes in condition.

In the past five years, small-scale sensors have emerged to address the above challenges. These commercially available sensors, although relatively small in size, have not been systematically evaluated, and many of them perform poorly relative to the reference methods. A few of their associated challenges are their low sensitivity, accuracy, drift, reliability, power, and cost. Commercially available low-cost optical particle counters based on light scattering cannot accurately measure particles smaller than 300 nm [13]. Counters based on complementary metal-oxide semiconductor (CMOS) image sensing/holographic detection combined with extensive image processing are restricted to detecting particle size above 3 μm [13]. Table 1 shows the comparative performance of the available technology for particle detection and concentration measurement.

**Table 1 Comparative chart amongst the existing technology for particle detection and concentration measurement**

| Parameters | Technology | | | | |
|---|---|---|---|---|---|
| | Optical[1] [11] | QCM[2] | SAW[2] | BAW/FBAR[1,2] | MEMS[1,2] |
| Mass concentration range | 1 to 1000 μg/m$^3$ | 2 to 2000 μg/m$^3$ [8] | 44 to 66 μg/m$^3$ [30] | N/A | 50 to 200 μg/m$^3$ [28] |
| Mass concentration resolution | 1 μg/m$^3$ | 6.5 μg/m$^3$ | 44 μg/m$^3$ [30] | 2 μg/m$^3$ [29] | 1 ng/m$^3$ [19] |
| Limit of detection | 0.3 μm | N/A | 2 μm | N/A | 12 nm [7] |
| Concentration measurement limit | N/A | 2 μg/m$^3$ | 40 μg/m$^3$ [30] 0.17 ng [31] | 2 μg/m$^3$ [32] | 0.7 μg/m$^3$ [18] 29.8 fg [28] |
| Average supply current | 55 mA | N/A | N/A | N/A | 150 mA[3] |
| Size | 40.6 × 40.6 × 12.2 mm$^3$ | 1200 × 500 × 500 mm$^3$ [33] | 50 × 15 × 3 mm$^3$ [30] | 22.5 mm/7 mm [7] | 4 × 4 × 0.55 mm$^3$ [7] |

[1]PM$_{1.0}$, PM$_{2.5}$, PM$_4$ and PM$_{10}$, [2]PM$_{2.5}$, [3]With integrated planer coil

## 3. MEMS sensors

MEMS sensors are promising candidates for the detection and measurement of several chemical/biological entities [13][34]. Due to their ultra-small size, MEMS devices offer the potential to precisely detect and quantify the size and number of particles. As observed in Table 1, miniaturization with MEMS helps to attain the limit of detection of about 29.8 *fg* and 12 *nm* for mass concentration and the particle size, respectively Moreover, MEMS technology offers advantages such as low power, size, weight, reliability, and eventually the scaling potential to reduce the development cost. MEMS sensors provide direct detection of particles rather than monitoring changes of parameters that could lead to the formation of particles. This allows the detection of even subtle changes of particle concentration. Moreover, MEMS sensors can provide a direct mass measurement, as opposed to reference measuring instruments, such as optical particle counters, which indirectly estimate mass [22][23][35].

### 3.1. Micro-resonant mass sensors

Amongst the available sensing mechanisms in MEMS, resonant sensing is a promising method for detecting minute parametric variations in the structural properties of a vibrating element. Representative examples of resonant sensing are mass detection and measurement [36–38], strain measurement [39–41], angle detection [42,43], and pressure sensing [44–48]. In resonant sensors, the output signal is the variation/shift in the resonant frequency ($\Delta f$) of a vibrating element that is subjected to small perturbations in structural parameters, typically effective stiffness/mass. Some advantageous features of resonant sensing are its simple mechanical design, quasi-digital nature of the signal, and ultra-high resolution [49–51] (up to $10^{-18}$ grams scale [52–55]). Accurate frequency references, such as crystal oscillators, offer better stability than typical voltage or current references. Moreover, the output signal frequency of the resonant sensor is considered to be relatively immune to noise and interference [56]. Figure 1 shows a representative schematic diagram of a micro-vibrating element (cantilever) used for the detection of bio-chemical quantities. For the actuation, a piezoelectric thin-film is integrated on the surface of a cantilever [57].

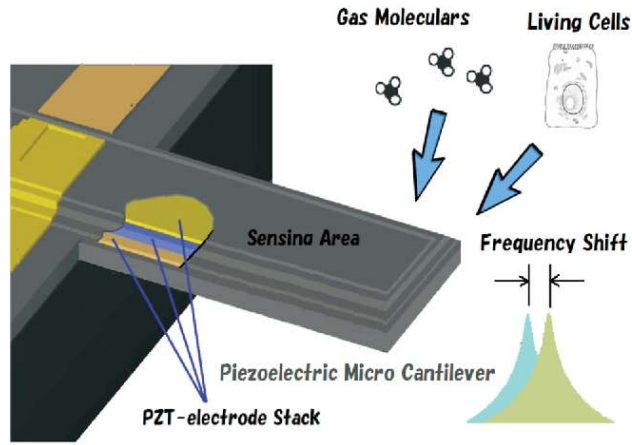

**Figure 1 Piezoelectric micro cantilever resonant bio-chemical sensor. The frequency shift is the output of the sensor. Reprinted from [57], with permission from JASE.**

The normalised output of a resonant mass sensor is given as

$$\frac{\Delta \omega_i}{\omega_i} = -\frac{1}{2}\frac{\Delta m}{M_{eff}}, \quad (1)$$

where $\Delta \omega_i$ is the shift in the resonant frequency in rad/s, $\omega_i$ is the resonant frequency in rad/s, $\Delta m$ is the change in mass, and $M_{eff}$ is the effective mass of the vibrating element. Therefore, sensitivity due to the added mass is

$$\left|\frac{\frac{\Delta \omega_i}{\omega_i}}{\frac{\Delta m}{M_{eff}}}\right| = \frac{1}{2}. \quad (2)$$

From equation (2), it can be inferred that for a given effective mass $M_{eff}$ and a resonant frequency $\omega_i$, $\Delta m$ can be quantified by measuring the frequency shift $\Delta \omega_i$. In the mass sensing experiments, it is assumed that the spring of the resonator remains unaltered while mass of the resonator undergoes a change. Another assumption taken into account is that the mass deposition is assumed to be uniform (as opposed to the case where distributed mass loading is shown [58]) on the surface of the resonator. Mass sensitivity has also been linked with the position of the added mass ($\Delta m$) on the surface of the resonator [59].

### 3.1.1. Piezoelectric transducer

As depicted in Figure 1, thin-film piezoelectric micro-resonant transducers have been widely used for particle mass detection and concentration measurements [17, 17, 21, 27, 29, 33, 58, 60, 61]. Piezoelectric resonant structure offers the following advantages. i) Relatively higher coupling coefficient (as opposed to the capacitive transduced resonators). This leads to the higher quality factor, $Q$, and lower motional impedance. ii) No DC bias requirement for actuation (as opposed to the electrostatically actuated resonator). iii) Improved compatibility with the CMOS fabrication process. The features i) and ii) greatly ease the requirements of 50 Ω impedance matching and the readout electronics, respectively. Figure 2 shows a representative schematic of a piezoelectric MEMS resonator. As shown in Figure 2(a), during fabrication, the piezoelectric layer (aluminium nitride (AlN)/zinc oxide (ZnO)/lead zirconate titanate (PZT)) is stacked on top of the silicon. The piezo layer is embedded between the two electrodes (for example, platinum (Pt) and/or gold (Au)). These electrodes are used for the charge collections. Silicon dioxide ($S_iO_2$) acts as an insulating layer. Figure 2(b) is the equivalent circuit model of this transducer, where $R$, $L$, and $C$ form the series resonant tank circuit. $C_p$ is the static capacitance of the piezoelectric stack, and $R_p$ models the parasitic leakage current of the piezo layer. From the model, the resonant frequency $f_r$ and $Q$ can be derived as $f_r = \frac{1}{2\pi\sqrt{LC}}$ and $Q = \frac{1}{R}\sqrt{\frac{L}{C}}$, respectively.

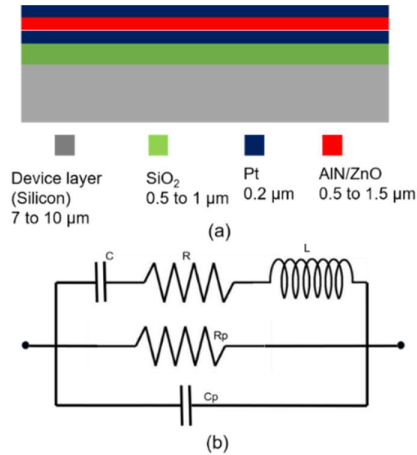

**Figure 2** Structure of a piezoelectric transducer and its equivalent electrical model.

### 3.1.2. Piezoelectric mass sensors

Quartz crystal microbalance (QCM) has been traditionally used in determining the mass sensitivity in many scientific applications [62]. A smaller diameter of electrode in QCM leads to higher mass sensitivity. The mass sensitivity of QCM is dependent on its resonant frequency, which, in turn, is mainly determined by the thickness of the quartz crystal as shown in the equation below [13]:

$$\Delta f = \frac{2 f_o^2}{\sqrt{\mu \rho}} \frac{\Delta_m}{A}. \quad (3)$$

In equation (3), $\Delta f$ is the shift in resonant frequency due to the added mass $\Delta_m$, $f_o$ is the fundamental resonant frequency, $\mu$ is the elastic constant, $\rho$ is the density, and $A$ is the surface area. A thinner piezoelectric crystal leads to a higher resonant frequency and, therefore, greater sensitivity. It is also reported that there exists an optimal diameter of the electrode beyond which mass sensitivity will drop in QCM [62]. The rationale behind this is that as the electrode diameter decreases, the motional resistance of the resonators will decrease, so the resonant amplitude will also decrease, which will ultimately result in the drop of the absolute mass sensitivity in QCM [62].

Previously, a QCM real-time oscillator was presented to detect 6.5 µg/m³ of PM$_{2.5}$ in the air [8]. A sensor detection limit of 2 µg/m³ was demonstrated. A high-end detection limit of about 2000 µg/m³ was recorded, showing the dynamic range of the sensor. Field tests showed the readings of the proposed QCM sensor were correlated with the readings obtained by the commercial air quality detector. Given the currently modified guidelines of WHO for PM$_{2.5}$ [1][12], this sensor was proposed to be used as a handheld, low-cost solution for personal exposure monitoring. The linear response of the sensor in the range -6 µg/m³ to 100 µg/m³ was proposed to cover the typical ambient PM$_{2.5}$ concentration range in urban areas. In another work, the response of the device to the mass loading of the particles showed a negative frequency shift, obtaining a sensor responsivity of approximately 220 kHz/ng [13]. Other acoustic wave based devices, such as surface/bulk acoustic wave (S/BAW) and thin-film bulk acoustic resonator (FBAR), have also been proposed to be used in particle detection and mass measurement (refer Table 1) [8, 13, 60, 62].

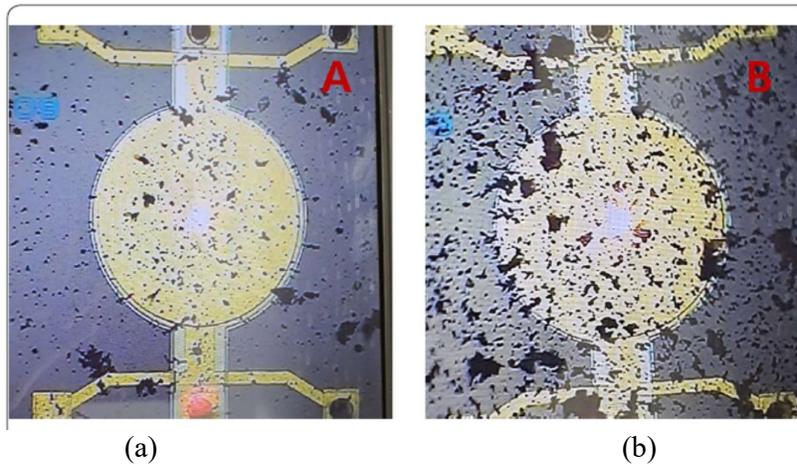

(a)                                           (b)

**Figure 3 Thin-film piezo MEMS resonator that was used to characterise mass deposition rate using frequency shift method, where (a) and (b) show smaller and**

**larger concentrations of the deposited mass, respectively. Reprinted from [63] with permission from Springer Open.**

A membrane based piezo-electric MEMS resonator was constructed and characterised for mass estimation [63]. The resonant frequency shift based sensitivity of the device was 0.354 kHz/pg. The lowest mass deposition of about 116 pg was estimated through the optical inspection. Figure 3 presents the microscopic view of the device, showing two cases: (a) a smaller concentration and (b) a larger concentration of adsorbed test particles. A MEMS piezoelectric thin-film resonator was used as an oscillatory mass balance to characterise the mass deposition in the PM measurement system [10]. The device was operated at the resonant mode frequency of 3.1 MHz and $Q$ of 2100, under atmospheric pressure and at room temperature. In this work, silver nanoparticles and poly-styrene latex nanoparticles were used to demonstrate the mass measurement under laboratory conditions. A normalised frequency shift ($\Delta f/f$) and mass change ($\Delta m$) were recorded with respect to time for the cases of i) slow and ii) accelerated nanoparticle deposition. The device sensitivity was characterised at different time intervals. The short-term frequency stability (16 ppb) of the MEMS oscillator based mass balance was correlated with the estimate of the minimum detectable mass (pg). By measuring the maximum deposition of the mass (i.e., 2.5 µg), the dynamic range of the sensor was characterised. In another study, silver particles from an inkjet printer were used to characterise the mass sensitivity of the sensor. The maximum deposition of mass of 8.89 ng was determined in the linear sensing range [33]. Mass concentrations (aerosol particles) in the range of 10 µg/m$^3$ to 1000 µg/m$^3$ were reported [6].

A piezoelectric MEMS resonator based oscillator was characterised for mass loading effect with the sensitivity of 8.8 Hz/ng [29]. The authors chose to operate the resonator in its higher mode of operation (at 1 MHz) to enhance mass sensitivity. A deposition of cigarette particles onto the surface of the resonator resulted in the decrease of the resonant frequency of the selected mode and also the resultant $Q$. A slight decrease in $Q$ was correlated to the increased viscous damping in the resonator due to the larger size of the particles deposited.

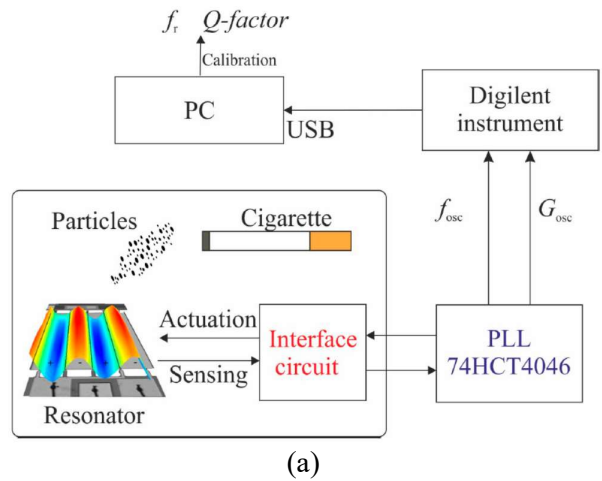

(a)

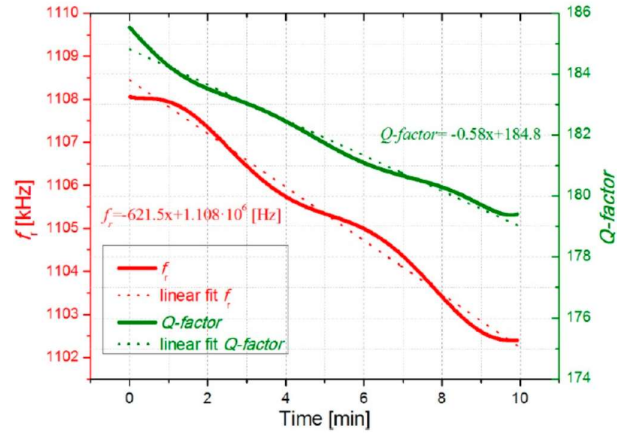

(b)

**Figure 4 (a) A piezoelectric MEMS oscillator for the characterisation of cigarette particle detection. (b) Output graph showing the variation in the measured resonant frequency and the device $Q$ as a function of time during mass-loading. Copyright © *2019* by MDPI. All rights reserved. Reprinted, with permission, from [29].**

Figure 4 shows the output graphs of the resonant oscillator. This graph shows the variation of the MEMS oscillator frequency and $Q$ with respect to time [29]. This implementation provided a relatively inferior detection limit (1.14 μg). However, this project used off-the shelf, low-cost electronic components, such as phase-lock-loop (PLL) integrated circuit (IC), and it demonstrated the real-time sensor output. This is opposed to the results produced using highly accurate test instruments, such as lock-in amplifiers, which show detection limits on the scale of picograms [10,33].

The frequency response of the MEMS piezoelectric AlN-on-Si weakly coupled resonators (WCR) was obtained by measuring i) shifts in the amplitude ratio (AR) and ii) shifts in the resonant frequency of the resonators [35]. These outputs were then correlated directly to the mass of the soot particles accumulated on the resonator surface. The long-term stability of this MEMS oscillator sensor was characterised for both of its viable readouts, i.e., AR and frequency shift. The soot particle concentration (~1000 particles/cm$^3$) was generated and was impacted onto the surface of one of the resonators. In WCR sensors, the sensitivity of AR and the frequency with respect to the applied mass perturbation, respectively, were derived as below:

$$S_{R_i} = \frac{\partial \left(\frac{x1}{x2}\right)_{ji}}{\partial (\delta_m)} = \frac{\left|\frac{\Delta m}{2K_c}\right|}{\left|\frac{\Delta m}{m_{eff}}\right|} = \left|\frac{\Delta m}{2K_c}\right| \times \left|\frac{m_{eff}}{\Delta m}\right| \approx \left|\frac{m_{eff}}{2K_c}\right|, \quad (4)$$

$$S_{f_i} = \frac{\partial (\omega_i)}{\partial (\delta_m)} = \frac{\left|\frac{\Delta m}{2m_{eff}}\right|}{\left|\frac{\Delta m}{m_{eff}}\right|} = \left|\frac{\Delta m}{2m_{eff}}\right| \times \left|\frac{m_{eff}}{\Delta m}\right| \approx \frac{1}{2}. \quad (5)$$

In equations (4) and (5), $\frac{\Delta m}{m_{eff}}$, $\frac{\Delta m}{2K_c}$, and $\frac{\Delta m}{2m_{eff}}$ are the normalised values (unit less) of the mass perturbation, AR, and frequency output, respectively. Additionally, $K_c$ is the spring constant (N/m) of a spring that couples the resonators [65]. By comparing the above equations, the relative gain in the AR sensitivity by a factor of $\left|\frac{m_{eff}}{K_c}\right|$ can be obtained. The slopes of the curves in Figure 5 (a), shows about 3 orders high AR sensitivity (relative) is possible with the WCR used for particle detection. Input referred stability (refer to Figure 5(b)) of the two output metrics, i.e., AR and frequency shift, was characterised to determine the minimum detectable mass. The estimated mass resolution for AR and the frequency shift output were 367.8 pg and 2.16 pg, respectively [35]. As noted, the mass resolution of the AR measurement was lower compared to the mass resolution based on the frequency measurement. However, AR was proposed to be a readout for its relatively better stability to dictate the ability of the WCR sensor to quantify mass accumulation over long-term measurements.

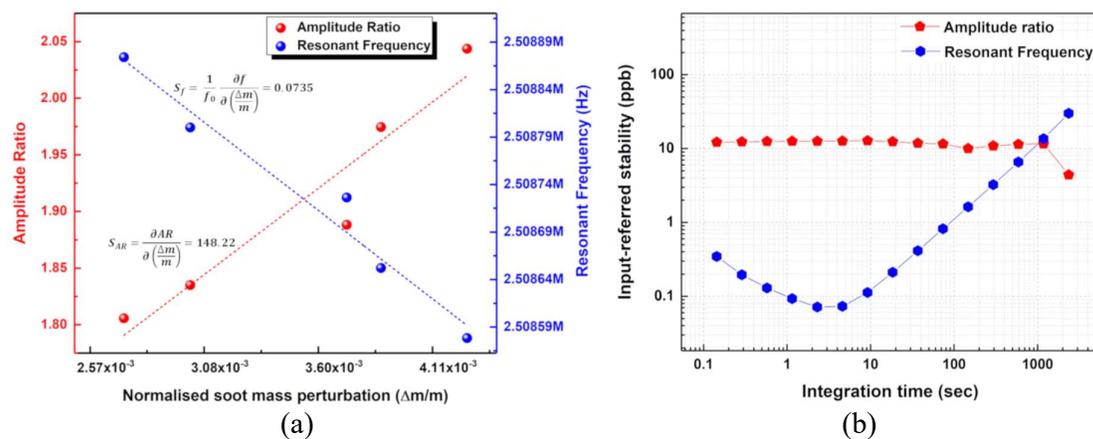

**Figure 5 (a) Sensitivity graphs of the coupled MEMS resonators based on AR shift and frequency shift when soot particles were impacted on one of the coupled resonators. (b) Estimated input referred stability using the sensitivity values indicating suitability of the AR for long-term measurements. Copyright © *2020* by MDPI. All rights reserved. Reprinted, with permission, from [35].**

### 3.1.3. Thermally actuated resonant transducer

Thermal-piezo-resistive transducers have also been used for the detection and concentration of ultra-fine PM [66][18][7]. In thermally actuated transducers, a current is passed through the microbeam. The current generates heat, and the temperature increase leads to displacement through thermal expansion. Concurrently, an AC current passing through the beam causes fluctuating temperature, and, in turn, alternating thermal stress in the beam. This resulting thermal stress actuates the resonator in its resonant mode. Variation in the resistance ($\Delta R/R$) in the same actuator beam due to the mechanical vibrations modulates the beam resistance. This piezo-resistive effect produces time-varying current, the frequency of which can be measured. Figure 6(a) shows the COMSOL simulated temperature field on the polysilicon microbeam surface, solving the model using a temperature-dependent resistivity. Based on the colour scale, the maximum temperature is about 506 K. Figure 6(b) shows the deformation of the microbeam. The displacement for the temperature-dependent case is 14 nm.

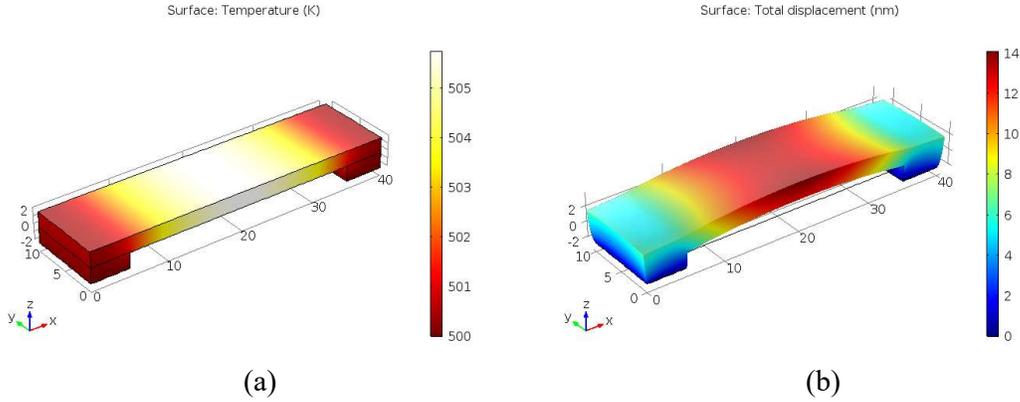

(a)                                          (b)

**Figure 6 COMSOL multiphysics models of (a) temperature distribution along the surface of a beam and (b) beam displacement due to the thermal expansion.**

A thermally actuated and piezo-resistively sensed silicon resonator in the megahertz scale was tested to determine aerosol particle mass in the air [67]. Shifts (downwards) in the resonant frequencies and the $Q$ of the sensor were recorded as a function of mass loading for over a period of 80 seconds. Air particle detection in the range of 100 ng to 1000 ng was reported. A work in [7], demonstrated a chip-scale implementation of a real-time, thermally actuated, and piezo-resistively sensed resonant mass balance with an operating frequency of about 5 MHz. The PM was measured in three different environments: a regular laboratory room, the outlet of an air purifier and a 10,000 class cleanroom, and the results were 202, 43, and 4 ng/m$^3$, respectively. These results showed that the mass sensitivity (frequency shift slope) in each environment decreased, meaning the frequency shift became slower when fewer particles were present in the sampled air (in the cleanroom). In the best case, the calculated sensitivity of the resonant air quality sensor was 42 Hz/pg [7]. To enable further integration, a chip-scale integration of the impactor and the sensor was demonstrated [7]. Such advancement offers a possibility for a handheld, relatively small MEMS-based air pollution monitor.

In another work, a mass concentration of carbon nanoparticles was measured via a cantilever based gravimetric sensor [18]. Two prototype sensors were used with the frequencies of about 89 kHz and 201 kHz. Electro-thermal actuation and piezo resistive detection were used, supplemented by the $Q = 2000$ (in-plane) and $Q = 189$ (out-of-plane motion), respectively. Notably, the limit of detection for the out-of-plane resonators with a $Q = 189$ offered an improved limit of detection of about 0.7 µg/m$^3$. Reported mass sensitivities were 0.013 Hz/pg and 0.14 Hz/pg. The equation used to calculate the mass concentration was given as $C_m = C_f \frac{\Delta f}{\Delta t}$. Here, $C_f$ is the calibration factor, and $\Delta f$ is the frequency shift in a particle collection period $\Delta t$. The limit of detection of mass concentration was determined as $\frac{3 \times C_f \times \sigma}{\Delta t}$, where $\sigma$ is the frequency noise floor obtained over a period $\Delta t$. Figure 7(a) shows a graph of i) the number of particles per cubic centimetre as per the size of the particles, and ii) the mass concentration (µg/m$^3$) as per the size of the particles. Figure 7(b) provides information about the total number of particles per cubic centimetre and the accumulated mass (µg/m$^3$) measured over a time period of about 6 hours. About 20,000–66,000 particles/cm$^3$, corresponding to mass concentrations of about 7–31 µg/m$^3$, were estimated through the measurement. In a work presented by [68], mass concentrations in the range <10 µg/m³ were shown with a limit of detection of ~1 µg/m³ within a particle collection time of ~10 min.

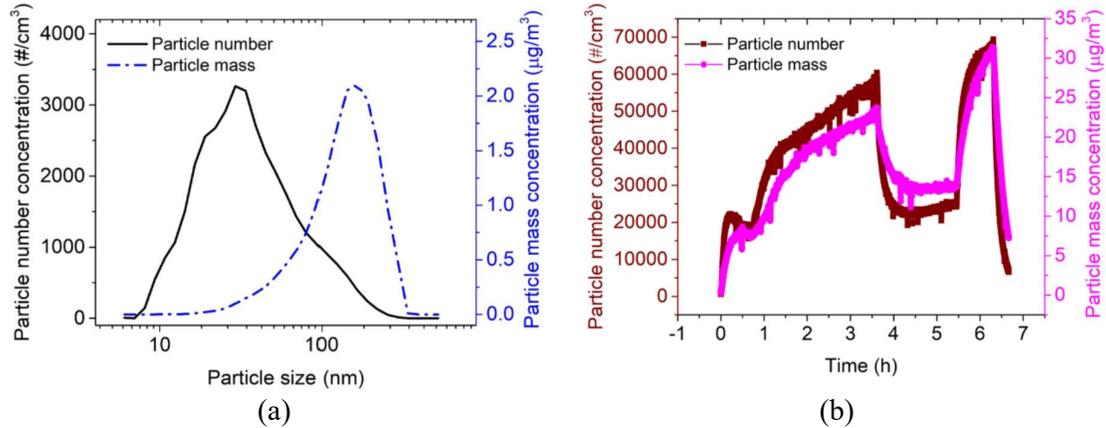

(a)                                                                 (b)

**Figure 7 Graphs showing (a) the number of particles and mass concentrations as per the particle size and (b) the number of particles and mass concentrations with respect to time measured by cantilever based gravimetric sensing platform. Copyright © *2020* by MDPI. All rights reserved. Reprinted, with permission, from [18].**

### 3.2. Particle detection and concentration measurement in high viscous media

For the resonator operating in the fluidic environment, the following equation can be used to quantify the fluidic mass :

$$\left(\frac{f_r}{f_o}\right)^2 = \frac{m_r}{m_T}\left(1-\frac{1}{2Q^2}\right). \quad (6)$$

In this equation, $f_r$ and $f_o$ are the resonant frequencies in the vacuum and in the fluid, respectively. The quantities $m_r$ and $m_T$ are the masses when the resonator operates in a vacuum and when it is submerged in the fluid to be analysed, respectively. Total mass is given as $m_T = m_r + m_{fl}$, where $m_{fl}$ is the effective mass of the displaced fluid during each oscillation cycle. With sufficiently high $Q$, the term in the parentheses used in the equation above can be neglected to approximately quantify the effective value of $m_{fl}$. Measurement in liquid is challenging as it drastically decreases the $Q$ and resonant frequency of the resonator. A solution is to use higher-order or less damped modes (in-plane) of the resonators [18, 29][59]. The higher order flexural mode of a cantilever, as shown in Figure 8, was used for the measurements in liquid [27]. This mode was used due to its low viscous damping.

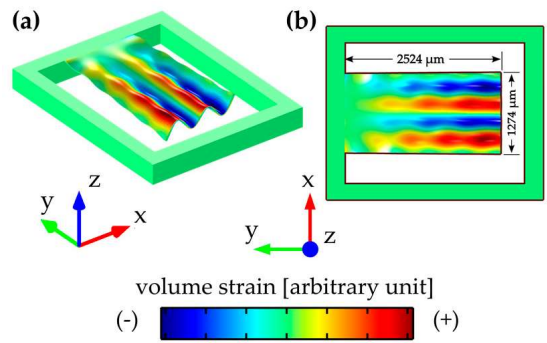

**Figure 8 Representative higher order (specifically, 4th order) vibration mode of a resonant cantilever used for particle detection in air** [29] **and in liquid** [61][27]**. This mode offers relatively higher *Q* (due to the lower viscous damping) for sensing in liquid. Copyright © *2017* by MDPI. All rights reserved. Reprinted, with permission, from** [27]**.**

The ferrous particles are considered to be responsible for the wear in the mechanical components used in the industry [61]. With a transducer as shown in Figure 8, the concentration of the debris (ferrous particles) in the lubricant oil was measured [17]. Figure 9(b) shows such measurement over time. Up to 500 ng of accumulated iron oxide ($Fe_3O_4$) nanoparticles was measured over a time period of 120 minutes (Figure 9(b)). The concentration of about 100 μl of highly viscous liquid (bitumen) [69] was deposited onto the surface of the resonator operating at 3.5 MHz resonant frequency [61]. Viscosity changes at varying temperature of a bitumen were measured by the resonator. The resonator exhibited a *Q* of 4 during the operation. The results were correlated with the standard test equipment used to measure a viscosity. It is to be noted that if the resonant sensor exhibits the low *Q* in the vacuum/air, it is challenging to obtain the sensor response in the liquids. Figure 9(a) shows the ethanol concentration measurements at different operating modes of the resonator. A high % of ethanol indicates the normal fermentation process, and lower % of ethanol prompts inadequate fermentation process. This study highlights the detection and measurement of the particles in remote monitoring of food quality [61].

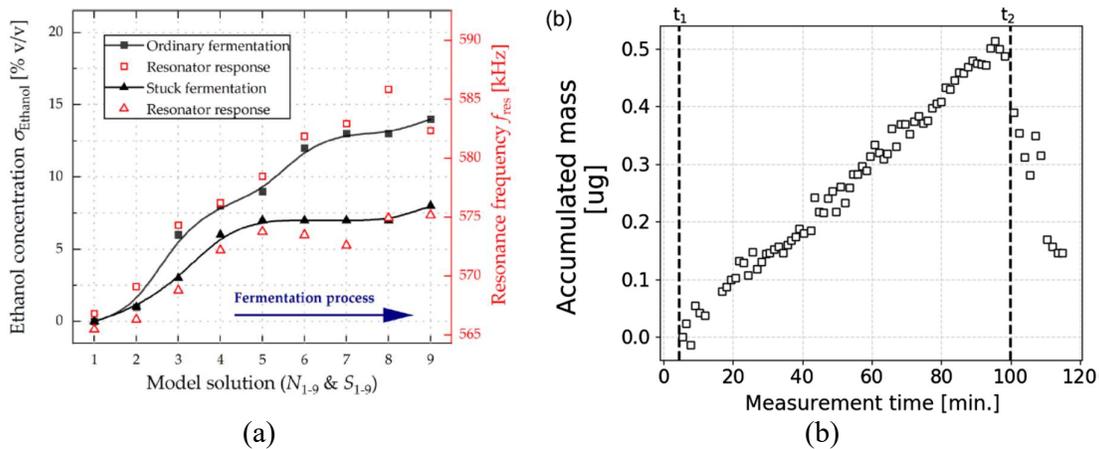

**Figure 9 (a) The ethanol concentration measurements at different operating modes of the resonator for the normal condition (high percentage of ethanol) and when the fermentation process is not proper (lower percentage of ethanol)** [27]**. (b) Mass (iron oxide ($Fe_3O_4$) nanoparticles) accumulation over a defined time. Smallest measured particle size is 250 nm** [61]**. Copyright © *2017* by MDPI. All rights reserved. Reprinted, with permission, from** [27]**.**

An aluminium nitride (AlN) piezo MEMS resonator, operating in its higher order mode with a *Q* of 366, was used to quantify the density, $\rho$, and viscosity, $\mu$, of several liquids (deionized water, ethanol, and isopropanol, etc.). The relationship between *Q*, the viscosity, and the density of liquids was represented as $Q = 1/(\rho\mu)^{½}$. This shows the ability of the sensor to directly correlate to these parameters. The highest mass sensitivity was found in the edge regions of the free end of the cantilever (refer to Figure 8), where the deposited particles receive high kinetic energy to manifest highest $\Delta f$ [59].

For measurement of the changes in fluid properties (e.g., viscosity and density of oil), a sensor will be fully immersed in the target fluid [26]. For instance, in one study, the cantilevers were fully immersed in 50 mL of oil. Four oil samples were temperature treated (heated at 150°C) for different durations of time. This caused the varying degree of the oxidative degradation in the oil. The principle of dielectrophoresis was used to accumulate the mass particles onto the sensor surface [26]. A sensor was used to measure different particle concentrations due to the oxidative degradation of the oil constituents. The measured resonance frequency shift showed a correlation to the degree of oil oxidation. The decrease in the resonant frequency monitored over the period of 24 hours was correlated to the larger mass concentration, i.e., the longer degree of the oil ageing process [26]. A degradation of the quality factor $Q$ was also observed to evaluate the state of the lubrication oils used in industrial machines [26]. It was claimed that the energy fraction needed to deform the accumulated particle layer was lost during each oscillation cycle. Therefore, a degradation (slight) of the $Q$ was correlated with the larger concentration of the particles accumulated onto the sensor surface.

### 3.3. Particle accumulation principles

In the context of particle sensing and concentration measurement, different techniques have been used for the accumulation and sampling of the particles. In [18], interdigitated electrodes were integrated on the cantilever surface to promote the accumulation of particles via electrophoresis. In [26], particles in liquid were collected due to dielectrophoresis (particles of interest were dielectric and were polarized in the presence of an external electric). In [18], the dielectrophoretic force exerted on particles in a stationary electric field was defined as $F_{DEP} \approx -\varepsilon_{fl} R^3 \nabla E^2$, where $\varepsilon_{fl}$ is the permittivity of the fluid, $R$ is the radius of the particle, and $\nabla E$ is the gradient of the stationary electric field. In [17], the magnetophoresis effect was used to attract ferrous particles in liquid (deionised water) to the resonant cantilever using an integrated magnetic coil. A magnetic force in this case was given by $F_M = (1/2)(\nabla_\chi V_p/\mu_o)(\nabla B^2)$. Here, $\nabla_\chi$ is the magnetic susceptibility of the particle relative to the fluid, $V_p$ is the particle volume, $\mu_0$ is the magnetic field constant, and $B$ is the magnetic field density. The term $B^2$ indicates that the magnetophoretic force is directed along the magnetic field gradient. A micro-fluidic channel [70], an air-flow micro-channel integrated with a cantilever [66], and single/multiple stages of micro-nozzles [33][7] have been proposed to sample/separate the particles and thus determine the size and number of particles. In [6], an impactor stage (for sorting and sampling the air particle distribution) was assembled on a printed circuit board (PCB). A particle collection efficiency up to 50% (for both $PM_{2.5}$ and $PM_1$) was achieved with the custom made airflow impactor implemented and integrated with the sensor. In [30], based on the thermophoresis effect, the mass loading from the deposited $PM_{2.5}$ was used to mark the shift in the SAW velocity. Accordingly, the change of the oscillation frequency was measured for $PM_{2.5}$ mass detection. These techniques are too sophisticated, however, for a real-time sensor.

Table 2 provides a comprehensive summary of the features and performance of micro-resonant sensors used for particle detection and particle concentration measurement. The majority of the resonators feature operating frequency past 1 MHz. Most of the resonators are based on piezoelectrics that offer a reasonable $Q$ in air and liquid. Thermally excited and piezo-resistively sensed silicon resonators had the highest $Q$ amongst all the resonators with a limit of mass detection in the scale of femtograms. The AlN piezoelectric on silicon resonator had the measured mass concentration range and particle size of 49 to 532 ng and 3.5 μm (mean diameter), respectively [26]. Another work used an AlN piezoelectric resonator to achieve the mass concentration and the particle size range of 0.49 μg and 250 to 185 μm, respectively [61][17].

**Table 2 Summary of performance parameters of MEMS sensors used for particle detection and mass concentration measurement**

| Ref. | Technology | Q | Circuit | Resonant frequency | Sensitivity | Detection limit | Particles detected/measured |
|---|---|---|---|---|---|---|---|
| [33] | Thin-film piezoelectric-on-silicon | 311 | Oscillator | 5.12 MHz | 4.96 Hz/pg | 0.15 pg | Silver particles |
| [71] | Thermal-piezo resistive | 4300 | Oscillator | 4.4 MHz | 1.22 Hz/pg | 83 fg | Printed droplets |
| [18] | Thermal-piezo resistive | N/A | Oscillator | 89 kHz and 209 kHz | 0.013 and 0.14 Hz/pg | 1.4 and 0.7 μg/m$^3$ | Carbon nanoparticles |
| [19] | Thermal-piezo resistive | N/A | Oscillator | 201 kHz | N/A | 1 ng/m$^3$ | Carbon, titania, silica, cigarette particles |
| [10] | Thin-film piezoelectric-on SOI | 2100 | Oscillator | 3.1 MHz | N/A | N/A | Aerosol particles |
| [61] | AlN piezoelectric resonator | 366* | Resonator | 3.5 MHz | N/A | N/A | Ethanol concentrations in fermentation of grape must in wine |
| [61][17] | AlN piezoelectric resonator | 366* | Resonator | 2.35 MHz | 22.85 Hz/pg 82.85 pg/ Hz | 4 ng | Iron oxide ($Fe_3O_4$) nanoparticles - ferrous particles in lubricant oil |
| [26] | AlN piezoelectric on silicon | 120* | Resonator | 2.25 MHz | 14.4 Hz/ng to 13 Hz/ng | 3.5 ng | Particles in lubricant oil |
| [61] | AlN piezoelectric resonator | 4* | Resonator | 3.5 MHz | N/A | N/A | Viscosity of bitumen |
| [29] | AlN piezoelectric resonator | 185 | Oscillator | 1 MHz | 8.8 Hz/ng | 114 pg | Cigarette particles |
| [8] | Quartz microbalance | N/A | Oscillator | ≈32 kHz | N/A | 2 μg/m$^3$ | $PM_{2.5}$ |
| [6] | Lead zirconate titanate piezoelectric resonator on silicon | 438 | Oscillator | 4.05 MHz | N/A | 0.71 pg | $PM_{2.5}$ and $PM_1$ |
| [35] | Piezoelectric AlN-on-Si | 1773 | Oscillator | 2.5 MHz | N/A | 100 nm | Diesel soot |
| [9] | Piezoelectric membrane | 300 | Resonator | ≈1 MHz | 1.4 kHz/0.01 μg | 10 nm | PM |
| [66] | Silicon | 415 | Resonator | 98 kHz | 0.22 Hz/pg | 345 pg | Polystyrene (PS) micro-particles |

Q is quality factor, *measured in viscous fluids

## 4. Discussion

The applications of the MEMS resonant mass sensor to monitor food quality, industrial processes, and environmental parameter (i.e., PM) are reported. Due to their extremely high-quality factor, excellent output stability over long-term and short-term measurements, and compatibility to integrate with the mainstream electronics, MEMS resonant sensors are widely being used in wireless communication, wearable electronics, medical diagnostic, robotics, etc. Owing to their highly precise output, MEMS resonant components are also extremely useful to detect and measure environmental, physical, biological, and chemical quantities with high parametric sensitivities. In this paper, some insights into the basics of MEMS resonant sensors (methods of transduction, materials used, and design techniques) are provided. This is followed by a comprehensive, critical analysis of the performance parameters of the MEMS resonant sensors used for the detection and measurement of fine and ultra-fine particles. The models of micro-resonant mass sensors that use piezoelectric and thermally actuated resonant transducers are explained. The operation of micro-resonant sensors in contaminated gas (air) and highly viscous liquids (deionized water, lubrication oil, and bitumen) is analysed [61].

### 4.1. Challenges

Although micro-resonant gravimetric sensors are being widely deployed for particle detection and concentration measurement with the features such as accuracy, limit of detection, dynamic range, repeatability (sensor re-use), and portability, there are limitations however. First, the maximum theoretical parametric sensitivity $\Delta f/f$ is limited to 0.5 [64][72,73]. Second, resonant sensors are prone to pressure and/or temperature fluctuations. Compensatory electronics or embedded differential measurement [18][26] are required to cancel the sensor output to the false perturbations. Furthermore, resonant sensors, when used as a mass sensor, are able to detect only one type of target particle at a time. This avoids the possibility of rapidly detecting and differentiating multiple particles in parallel [74]. Long-term frequency stability appears to be the focus of ongoing research (where the output stability of the amplitude output and the frequency output of a resonant sensor are compared) [75].

From the design and fabrication point of view, piezoelectric transduction needs integration of piezoelectric thin film with the silicon. Such integration usually results in a drastic decrease in the resonator $Q$. On the other hand, higher $Q$ (possible with in-plane motion of thermally actuated resonators [67]) offers an improved mass resolution (limit of detection). Furthermore, piezoelectric resonators as reported in this paper mostly are operating with the frequencies past MHz. This overstrains the design efforts of sustaining electronics in sensing applications. Finally, while miniaturisation in MEMS does benefit to attain sensor performance, such as the lowest possible detection limit, it may also lead to a relatively lower transduction area, and therefore lower signal-to-noise (S/N) ratio.

### 4.2. Recommendations

The sensor performance can be further improved to sustain the current implementation efforts. The following suggestions are provided. a) The mass sensitivity of the resonant sensor could be further improved by the accumulation of particles to regions of higher vibration amplitudes [26]. b) Using resonant modes that support reduced viscous damping—for instance, in-plane motion of the resonators—is recommended for the mass sensor operating in highly viscous media (liquid) [59]. Alternately, a higher order mode was used in [29]. c) Adhesion of the sensing layers (e.g., polymers or nanostructures) onto the surface of a resonator has been linked with the enhancement of the sensor performance, such as reaching the ultimate detection limit.

### 4.3. Sensor repeatability

We predict the following challenges about the repeatability of the sensor. First, a mechanism will be needed to clean the resonator surface after each deposition of mass is quantified [9]. For instance, in [61] the sensor was rinsed with the fresh deionized water, and in [26] the sensor was rinsed thoroughly with heptane for several minutes, followed by drying in air. In [19], a sensor with electrostatic precipitator was used to remove deposited nanoparticles from the surface of the resonator so the monitor could be refreshed without disassembling. In [66], the repeated detection capability of the sensor was ensured by reversely pumping the trapped particles out of the micro-channel. Residual particles can inhibit the accumulation of new particles on the surface, and a possible solution proposed in [26] was to reduce surface adhesion by depositing a passivation layer (fluoropolymer) to cover the electrodes used to attract the particles. A few other challenges we foresee include real-time data measurement, as only a few of the studies covered in this review demonstrated this capability [6,18,33,33]; and point-of-requirement, referring to establishing the sensor network in areas not easily accessible. Nonetheless, the developed devices show the sensitivity of up to 22.85 Hz/pg [61][17], large, linear dynamic range (about 100 *pg* to 2.5 *μg*), the measurement of the mass size below 300 *nm* [19], and mass concentration in the *fg* level [26][32].

Developing a more desirable cleaning mechanism can facilitate the translation of the MEMS based sensing system to a commercially available handheld solution in the market. MEMS resonant sensor are therefore a popular choice to develop highly accurate, low-cost, reliable, portable, real-time, integrated sensing solutions for online monitoring of health, food quality, agriculture, industrial processes and range of environmental parameters to democratize sensing across the globe.

## 5. Conclusion

This scientific review reports the state-of-the-art in MEMS resonant sensors used for the detection and concentration measurement of particles. The fine and ultra-fine particles characterised are the nanoparticles, PM and ferrous particles. Given the detrimental impact of these particles on the human respiratory health, industrial machines, food, water quality, and many more, detection and concentration measurement of these quantities are extremely important. Measurement of these particles in air and high viscous fluids is reported. Particle detection range, size limit and the mass concentration limit are given. Relevant sensor applications include real-time air quality monitoring, embedded sensors for predictive machine maintenance, and food quality. This work identifies the future challenges and opportunities in the field. This will sustain the design effort to improve the sensor performance and/or meet the target specifications in the areas mentioned.

**Funding**

This research did not receive any specific grant from funding agencies in the public, commercial, or not-for-profit sectors.

**CRediT author statement**

**Vinayak Pachkawade:** Writing- Original draft preparation, Investigation and Editing, **Zion Tse:** Reviewing and Editing

**References**

[1]   Air pollution, (n.d.). https://www.who.int/health-topics/air-pollution#tab=tab_1 (accessed December 3, 2021).

[2]   WHO global air quality guidelines: particulate matter (PM2.5 and PM10), ozone, nitrogen dioxide, sulfur dioxide and carbon monoxide, (n.d.). https://www.who.int/publications/i/item/9789240034228 (accessed December 3, 2021).

[3]   WHO, Particulate matter (PM2.5 and PM10), ozone, nitrogen dioxide, sulfur dioxide

and carbon monoxide. Geneva: World Health Organization; 2021. Licence: CC BY-NC-SA 3.0 IGO., (2021).

[4]   New WHO Global Air Quality Guidelines aim to save millions of lives from air pollution, (n.d.). https://www.who.int/news/item/22-09-2021-new-who-global-air-quality-guidelines-aim-to-save-millions-of-lives-from-air-pollution (accessed December 3, 2021).

[5]   A. Schieweck, E. Uhde, T. Salthammer, L.C. Salthammer, L. Morawska, M. Mazaheri, P. Kumar, Smart homes and the control of indoor air quality, Renew. Sustain. Energy Rev. 94 (2018) 705–718. doi:10.1016/J.RSER.2018.05.057.

[6]   C.H. Weng, G. Pillai, S.S. Li, A PM2.5 Sensor Module Based on a TPoS MEMS Oscillator and an Aerosol Impactor, IEEE Sens. J. 20 (2020) 14722–14731. doi:10.1109/JSEN.2020.3010283.

[7]   M. Maldonado-Garcia, V. Kumar, J.C. Wilson, S. Pourkamali, Chip-Scale Implementation and Cascade Assembly of Particulate Matter Collectors with Embedded Resonant Mass Balances, IEEE Sens. J. 17 (2017) 1617–1625. doi:10.1109/JSEN.2016.2638964.

[8]   X. Qin, X. Xian, Y. Deng, D. Wang, F. Tsow, E. Forzani, N. Tao, Micro Quartz Tuning Fork-Based PM 2.5 Sensor for Personal Exposure Monitoring, IEEE Sens. J. 19 (2019) 2482–2489. doi:10.1109/JSEN.2018.2886888.

[9]   N. Singh, M.Y. Elsayed, M.N. El-Gamal, Realizing a Highly Compact Particulate Matter Sensor with a MEMS-Based Resonant Membrane, Proc. IEEE Sensors. 2019-Octob (2019) 1–4. doi:10.1109/SENSORS43011.2019.8956505.

[10]  M. Chellasivalingam, L. Somappa, A.M. Boies, M.S. Baghini, A.A. Seshia, MEMS Based Gravimetric Sensor for the Detection of Ultra-Fine Aerosol Particles, Proc. IEEE Sensors. 2020-Octob (2020). doi:10.1109/SENSORS47125.2020.9278870.

[11]  Indoor Air Quality | Sensirion, (n.d.). https://www.sensirion.com/en/environmental-sensors/indoor-air-quality/ (accessed December 8, 2021).

[12]  Ambient (outdoor) air pollution, (n.d.). https://www.who.int/news-room/fact-sheets/detail/ambient-(outdoor)-air-quality-and-health (accessed October 20, 2021).

[13]  C.Y. Huang, P.H. Chung, J.Z. Shyu, Y.H. Ho, C.H. Wu, M.C. Lee, M.J. Wu, Evaluation and selection of materials for particulate matter MEMS sensors by using hybrid MCDM methods, Sustain. 10 (2018) 1–35. doi:10.3390/su10103451.

[14]  X. Zhu, C. Zhong, J. Zhe, Lubricating oil conditioning sensors for online machine health monitoring – A review, Tribol. Int. 109 (2017) 473–484. doi:10.1016/J.TRIBOINT.2017.01.015.

[15]  Standard Test Method for Oxidation Characteristics of Inhibited Mineral Oils, (n.d.). https://www.astm.org/d0943-20.html (accessed December 8, 2021).

[16]  Standard Test Method for Oxidation Stability of Steam Turbine Oils by Rotating Pressure Vessel, (n.d.). https://www.astm.org/d2272-14a.html (accessed December 8, 2021).

[17]  F. Patocka, M. Schlögl, C. Schneidhofer, N. Dörr, M. Schneider, U. Schmid, Piezoelectrically excited MEMS sensor with integrated planar coil for the detection of

ferrous particles in liquids, Sensors Actuators B Chem. 299 (2019) 126957. doi:10.1016/J.SNB.2019.126957.

[18] A. Setiono, M. Bertke, W.O. Nyang'au, J. Xu, M. Fahrbach, I. Kirsch, E. Uhde, A. Deutschinger, E.J. Fantner, C.H. Schwalb, H.S. Wasisto, E. Peiner, In-Plane and Out-of-Plane MEMS Piezoresistive Cantilever Sensors for Nanoparticle Mass Detection, Sensors. 20 (2020) 618. doi:10.3390/s20030618.

[19] M. Bertke, W. Wu, H.S. Wasisto, E. Uhde, E. Peiner, SIZE-SELECTIVE ELECTROSTATIC SAMPLING AND REMOVAL OF NANOPARTICLES ON SILICON CANTILEVER SENSORS FOR AIR-QUALITY MONITORING Braunschweig University of Technology and LENA Braunschweig , GERMANY and Fraunhofer-WKI , Material Analysis and Indoor Chemistry , 1 (2017) 1493–1496.

[20] 1405-F TEOM$^{TM}$ Continuous Ambient Air Monitor, (n.d.). https://www.thermofisher.com/order/catalog/product/TEOM1405F (accessed November 29, 2021).

[21] P.H. McMurry, Chapter 17 A review of atmospheric aerosol measurements, Dev. Environ. Sci. 1 (2002) 443–517. doi:10.1016/S1474-8177(02)80020-1.

[22] G.O. Avendano, J.C. Dela Cruz, A.H. Ballado, L.G.R. Ulyzes, A.C.P. Atienza, B.J.G. Regala, R.C. Uy, Microcontroller and app-based air quality monitoring system for particulate matter 2.5 (PM2.5) and particulate matter 1 (PM1), HNICEM 2017 - 9th Int. Conf. Humanoid, Nanotechnology, Inf. Technol. Commun. Control. Environ. Manag. 2018-Janua (2018) 1–4. doi:10.1109/HNICEM.2017.8269517.

[23] C.R. Li, C.N. Hsu, Y.C. Lin, M.W. Hung, C.C. Yang, H.Y. Tsai, Y.J. Chang, K.C. Huang, W.T. Hsiao, Integrating temperature, humidity, and optical aerosol sensors for a wireless module for three-dimensional space monitoring, 2018 IEEE Sensors Appl. Symp. SAS 2018 - Proc. 2018-Janua (2018) 1–4. doi:10.1109/SAS.2018.8336731.

[24] P. Arroyo, J. Gómez-Suárez, J.I. Suárez, J. Lozano, Low-Cost Air Quality Measurement System Based on Electrochemical and PM Sensors with Cloud Connection, Sensors. 21 (2021) 6228. doi:10.3390/s21186228.

[25] P. Peterson, A. Aujla, K. Grant, A. Brundle, M. Thompson, J. Vande Hey, R. Leigh, Practical Use of Metal Oxide Semiconductor Gas Sensors for Measuring Nitrogen Dioxide and Ozone in Urban Environments, Sensors. 17 (2017) 1653. doi:10.3390/s17071653.

[26] F. Patocka, C. Schneidhofer, N. Dörr, M. Schneider, U. Schmid, Novel resonant MEMS sensor for the detection of particles with dielectric properties in aged lubricating oils, Sensors Actuators, A Phys. 315 (2020). doi:10.1016/j.sna.2020.112290.

[27] G. Pfusterschmied, J. Toledo, M. Kucera, W. Steindl, S. Zemann, V. Ruiz-Díez, M. Schneider, A. Bittner, J.L. Sanchez-Rojas, U. Schmid, Potential of Piezoelectric MEMS Resonators for Grape Must Fermentation Monitoring, Micromachines 2017, Vol. 8, Page 200. 8 (2017) 200. doi:10.3390/MI8070200.

[28] T.X.C. Corporation, Gated Cmos-Mems Thermal-Piezoresistive Oscillator-Based Pm2 . 5 Sensor With Enhanced Particle, (2018) 75–78.

[29] J. Toledo, V. Ruiz-Díez, M. Bertke, H.S. Wasisto, E. Peiner, J.L. Sánchez-Rojas,

Piezoelectric MEMS resonators for cigarette particle detection, Micromachines. 10 (2019) 1–13. doi:10.3390/mi10020145.

[30] J. Liu, W. Hao, M. Liu, Y. Liang, S. He, A novel particulate matter 2.5 sensor based on surface acoustic wave technology, Appl. Sci. 8 (2018). doi:10.3390/app8010082.

[31] W.C. Hao, J.L. Liu, M.H. Liu, S.T. He, Development of a new surface acoustic wave based PM2.5 monitor, Proc. 2014 Symp. Piezoelectricity, Acoust. Waves Device Appl. SPAWDA 2014. (2014) 52–55. doi:10.1109/SPAWDA.2014.6998524.

[32] I. Paprotny, F. Doering, P.A. Solomon, R.M. White, L.A. Gundel, Microfabricated air-microfluidic sensor for personal monitoring of airborne particulate matter: Design, fabrication, and experimental results, Sensors Actuators A Phys. 201 (2013) 506–516. doi:10.1016/J.SNA.2012.12.026.

[33] C.H. Weng, G. Pillai, S.S. Li, A Thin-Film Piezoelectric-on-Silicon MEMS Oscillator for Mass Sensing Applications, IEEE Sens. J. 20 (2020) 7001–7009. doi:10.1109/JSEN.2020.2979283.

[34] Y.-C. Lee, M.-L. Hsieh, P.-S. Lin, C.-H. Yang, S.-K. Yeh, T.T. Do, W. Fang, CMOS-MEMS technologies for the applications of environment sensors and environment sensing hubs, J. Micromechanics Microengineering. 31 (2021) 074004. doi:10.1088/1361-6439/ac0514.

[35] M. Chellasivalingam, H. Imran, M. Pandit, A.M. Boies, A.A. Seshia, Weakly coupled piezoelectric MEMS resonators for aerosol sensing, Sensors (Switzerland). 20 (2020). doi:10.3390/s20113162.

[36] K. Park, N. Kim, D.T. Morisette, N.R. Aluru, R. Bashir, Resonant MEMS mass sensors for measurement of microdroplet evaporation, J. Microelectromechanical Syst. (2012). doi:10.1109/JMEMS.2012.2189359.

[37] J.E.Y. Lee, B. Bahreyni, Y. Zhu, A.A. Seshia, Ultrasensitive mass balance based on a bulk acoustic mode single-crystal silicon resonator, Appl. Phys. Lett. (2007). doi:10.1063/1.2822405.

[38] L. Li, Simulation of Mass Sensor Based on Luminescence of Micro/Nano Electromechanical Resonator, IEEE Electron Device Lett. 38 (2017) 395–398. doi:10.1109/LED.2017.2661261.

[39] L. Belsito, M. Ferri, F. Mancarella, L. Masini, J. Yan, A.A. Seshia, K. Soga, A. Roncaglia, Fabrication of high-resolution strain sensors based on wafer-level vacuum packaged MEMS resonators, Sensors Actuators, A Phys. (2016). doi:10.1016/j.sna.2016.01.006.

[40] M. Crescentini, C. Tamburini, L. Belsito, A. Romani, A. Roncaglia, M. Tartagni, Ultra-Low Power CMOS Readout for Resonant MEMS Strain Sensors, Proceedings. (2018). doi:10.3390/proceedings2130973.

[41] S. Ghosh, J.E.Y. Lee, Resonant tuning fork strain gauge operating in air with decoupled piezoelectric transducers, in: Proc. IEEE Sensors, 2017. doi:10.1109/ICSENS.2017.8234100.

[42] M. Saukoski, System and Circuit Design for a Capacitive MEMS Gyroscope, 2008.

[43] B. Xiong, L. Che, Y. Wang, A novel bulk micromachined gyroscope with slots

structure working at atmosphere, Sensors Actuators, A Phys. (2003). doi:10.1016/S0924-4247(03)00296-6.

[44] S. Bianco, M. Cocuzza, S. Ferrero, E. Giuri, G. Piacenza, C.F. Pirri, A. Ricci, L. Scaltrito, D. Bich, A. Merialdo, P. Schina, R. Correale, Silicon resonant microcantilevers for absolute pressure measurement, J. Vac. Sci. Technol. B Microelectron. Nanom. Struct. (2006). doi:10.1116/1.2214698.

[45] H.S. Wasisto, S. Merzsch, A. Waag, E. Uhde, T. Salthammer, E. Peiner, Airborne engineered nanoparticle mass sensor based on a silicon resonant cantilever, Sensors Actuators, B Chem. (2013). doi:10.1016/j.snb.2012.04.003.

[46] S. Ren, W. Yuan, D. Qiao, J. Deng, X. Sun, Pressure sensor with integrated resonator operating at atmospheric pressure, Sensors (Switzerland). (2013). doi:10.3390/s131217006.

[47] M. Esashi, S. Sugiyama, K. Ikeda, Y. Wang, H. Miyashita, Vacuum-sealed silicon micromachined pressure sensors, Proc. IEEE. (1998). doi:10.1109/5.704268.

[48] C.J. Welham, J.W. Gardner, J. Greenwood, A laterally driven micromachined resonant pressure sensor, Sensors Actuators, A Phys. (1996). doi:10.1016/0924-4247(96)80130-0.

[49] D. Jin, J. Liu, X. Li, M. Liu, G. Zuo, Y. Wang, H. Yu, X. Ge, Tens femtogram resoluble piezoresistive cantilever sensors with optimized high-mode resonance excitation, in: Proc. 1st IEEE Int. Conf. Nano Micro Eng. Mol. Syst. 1st IEEE-NEMS, 2006. doi:10.1109/NEMS.2006.334906.

[50] X. Zou, P. Thiruvenkatanathan, A.A. Seshia, A high-resolution micro-electro-mechanical resonant tilt sensor, Sensors Actuators, A Phys. (2014). doi:10.1016/j.sna.2014.10.004.

[51] X. Zou, P. Thiruvenkatanathan, A.A. Seshia, Micro-electro-mechanical resonant tilt sensor with 250 nano-radian resolution, in: 2013 Jt. Eur. Freq. Time Forum Int. Freq. Control Symp. EFTF/IFC 2013, 2013. doi:10.1109/EFTF-IFC.2013.6702229.

[52] I.B. Baek, S. Byun, B.K. Lee, J.H. Ryu, Y. Kim, Y.S. Yoon, W.I. Jang, S. Lee, H.Y. Yu, Attogram mass sensing based on silicon microbeam resonators, Sci. Rep. (2017). doi:10.1038/srep46660.

[53] Z.J. Davis, A. Boisen, Aluminum nanocantilevers for high sensitivity mass sensors, Appl. Phys. Lett. (2005). doi:10.1063/1.1984092.

[54] K.L. Ekinci, X.M.H.H. Huang, M.L. Roukes, Ultrasensitive nanoelectromechanical mass detection, Appl. Phys. Lett. (2004). doi:10.1063/1.1755417.

[55] T. Ono, M. Esashi, Magnetic force and optical force sensing with ultrathin silicon resonator, Rev. Sci. Instrum. (2003). doi:10.1063/1.1623627.

[56] R. Abdolvand, B. Bahreyni, J.E.Y. Lee, F. Nabki, Micromachined resonators: A review, Micromachines. (2016). doi:10.3390/mi7090160.

[57] J. Lu, L. Zhang, H. Takagi, T. Itoh, R. Maeda, Hybrid piezoelectric mems resonators for application in bio-chemical sensing, J. Appl. Sci. Eng. 17 (2014). doi:10.6180/jase.2014.17.1.03.

[58] P. Joshi, S. Kumar, V.K. Jain, J. Akhtar, J. Singh, Distributed MEMS mass-sensor based on piezoelectric resonant micro-cantilevers, J. Microelectromechanical Syst. 28 (2019) 382–389. doi:10.1109/JMEMS.2019.2908879.

[59] F. Patocka, M. Schneider, N. Dörr, C. Schneidhofer, U. Schmid, Position-dependent mass responsivity of silicon MEMS cantilevers excited in the fundamental, two-dimensional roof tile-shaped mode, J. Micromechanics Microengineering. 29 (2019). doi:10.1088/1361-6439/ab062a.

[60] F.H. Villa-López, G. Rughoobur, S. Thomas, A.J. Flewitt, M. Cole, J.W. Gardner, Design and modelling of solidly mounted resonators for low-cost particle sensing, Meas. Sci. Technol. 27 (2016) 025101. doi:10.1088/0957-0233/27/2/025101.

[61] M. Schneider, G. Pfusterschmied, F. Patocka, U. Schmid, High performance piezoelectric AlN MEMS resonators for precise sensing in liquids, Elektrotechnik Und Informationstechnik. 137 (2020) 121–127. doi:10.1007/s00502-020-00794-w.

[62] X.H. Huang, W. Pan, J.G. Hu, Q.S. Bai, The Exploration and Confirmation of the Maximum Mass Sensitivity of Quartz Crystal Microbalance, IEEE Trans. Ultrason. Ferroelectr. Freq. Control. 65 (2018) 1888–1892. doi:10.1109/TUFFC.2018.2860597.

[63] J.S. Choi, W.T. Park, MEMS particle sensor based on resonant frequency shifting, Micro Nano Syst. Lett. 8 (2020) 4–9. doi:10.1186/s40486-020-00118-9.

[64] V. Pachkawade, State-of-the-Art in Mode-Localized MEMS Coupled Resonant Sensors: A Comprehensive Review, IEEE Sens. J. 21 (2021) 8751–8779. doi:10.1109/JSEN.2021.3051240.

[65] V. Pachkawade, State-of-the-art in Mode-Localized MEMS Coupled Resonant Sensors: A Comprehensive Review, IEEE Sens. J. (2021). doi:10.1109/JSEN.2021.3051240.

[66] Y. Bao, S. Cai, H. Yu, T. Xu, P. Xu, X. Li, A resonant cantilever based particle sensor with particle-size selection function, J. Micromechanics Microengineering. 28 (2018) 085019. doi:10.1088/1361-6439/aabdac.

[67] A. Hajjam, A. Rahafrooz, J.C. Wilson, S. Pourkamali, Thermally actuated MEMS resonant sensors for mass measurement of micro/nanoscale aerosol particles, Proc. IEEE Sensors. (2009) 707–710. doi:10.1109/ICSENS.2009.5398557.

[68] M. Bertke, J. Xu, A. Setiono, I. Kirsch, E. Uhde, E. Peiner, Fabrication of a microcantilever-based aerosol detector with integrated electrostatic on-chip ultrafine particle separation and collection, J. Micromechanics Microengineering. 30 (2020) 014001. doi:10.1088/1361-6439/ab4e56.

[69] J. Toledo, V. Ruiz-Díez, G. Pfusterschmied, U. Schmid, J.L. Sánchez-Rojas, Calibration procedure for piezoelectric MEMS resonators to determine simultaneously density and viscosity of liquids, Microsyst. Technol. 24 (2018) 1423–1431. doi:10.1007/s00542-017-3536-0.

[70] D.P. Poenar, Microfluidic and micromachined/MEMS devices for separation, discrimination and detection of airborne particles for pollution monitoring, Micromachines. 10 (2019). doi:10.3390/mi10070483.

[71] C.C. Chu, S. Dey, T.Y. Liu, C.C. Chen, S.S. Li, Thermal-Piezoresistive SOI-MEMS Oscillators Based on a Fully Differential Mechanically Coupled Resonator Array for

Mass Sensing Applications, J. Microelectromechanical Syst. 27 (2018) 59–72. doi:10.1109/JMEMS.2017.2778307.

[72] C. Zhao, M.H. Montaseri, G.S. Wood, S.H. Pu, A.A. Seshia, M. Kraft, A review on coupled MEMS resonators for sensing applications utilizing mode localization, Sensors Actuators, A Phys. 249 (2016) 93–111. doi:10.1016/j.sna.2016.07.015.

[73] F.J. Giessibl, A direct method to calculate tip-sample forces from frequency shifts in frequency-modulation atomic force microscopy, Appl. Phys. Lett. (2001). doi:10.1063/1.1335546.

[74] B.E. DeMartini, J.F. Rhoads, S.W. Shaw, K.L. Turner, A single input-single output mass sensor based on a coupled array of microresonators, Sensors Actuators, A Phys. (2007). doi:10.1016/j.sna.2007.02.011.

[75] M. Pandit, C. Zhao, G. Sobreviela, A. Mustafazade, S. Du, X. Zou, A.A. Seshia, Closed-Loop Characterization of Noise and Stability in a Mode-Localized Resonant MEMS Sensor, IEEE Trans. Ultrason. Ferroelectr. Freq. Control. (2019). doi:10.1109/TUFFC.2018.2878241.